\documentclass{INTERSPEECH2023}

\interspeechcameraready

\title{SpeechBlender: Speech Augmentation Framework \\ for Mispronunciation Data Generation}
\name{Yassine El Kheir, Shammur Absar Chowdhury$^{*}$\thanks{$^{*}$ Corresponding author}, Ahmed Ali, Hamdy Mubarak, Shazia Afzal}

\address{Qatar Computing Research Institute, HBKU, Doha, Qatar
}
\email{\{yelkheir,shchowdhury,amali,hmubarak,safzal\}@hbku.edu.qa}

\begin{document}

\maketitle
\begin{abstract}

The lack of labeled second language (L$2$) speech data is a major challenge in designing mispronunciation detection models. 
We introduce SpeechBlender - a fine-grained data augmentation pipeline for generating mispronunciation errors to overcome such data scarcity. The SpeechBlender utilizes varieties of masks to target different regions of phonetic units and use the mixing factors to linearly interpolate raw speech signals while augmenting pronunciation. The masks facilitate smooth blending of the signals, generating more effective samples than the `Cut/Paste' method. Our proposed technique showcases significant improvement at the phoneme level in two L2 datasets, we achieved state-of-the-art results on ASR-dependent mispronunciation models with publicly available English Speechocean762 testset, resulting in a notable $5.0$\% gain in Pearson Correlation Coefficient (PCC). Additionally, we benchmark and demonstrate a substantial $4.6$\% increase in F1-score with the Arabic AraVoiceL2 testset.

\end{abstract}

\section{Introduction} 
\label{sec:intro}

With the advent of globalization, more people are inclined towards multilingualism. Such a shift in interest promotes the demand for the computer-aided pronunciation training (CAPT) system to language learning applications \cite{CAPT,CAPT_2}. The CAPT system potentially exploits the advancement of speech technology along with curriculum management and learner assessment among others. Mispronunciation detection (MD) and diagnosis system -- a crucial component of the CAPT -- detects the mispronunciation in an L2 learner‘s speech and highlights the error to give effective feedback. 
Researchers studied many methods to assess pronunciation quality. A majority of these methods use pre-trained automatic speech recognition (ASR) systems to either \textit{(i)} pinpoint the mismatch between the aligned ASR output sequence with the reference; or \textit{(ii)} utilize the log-posterior probability from the ASR to compute variants of the goodness of pronunciation (GOP) scores \cite{GOP_WITT,GOP_1,GOP_2} to asses pronunciation.
One successful line of research treated the mispronunciation detection as a regression, or a classification task \cite{transfer_}. Using deep learning, the network is either trained in an end-to-end manner \cite{non_ASR,non_ASR_2,non_ASR_3,DNN_3} or built upon the computed GOP features from the pre-trained ASR \cite{JIM, likeJIM_asr, 3M} to detect errors. These approaches are heavily dependent on the availability of a large amount of manually annotated datasets for the target demographic L2 speakers. However, such annotated datasets are limited and often contain uneven positive class labels.\footnote{In the case of MD, the positive class represents -- missed, mispronounced, or accented phonemes, and the negative classes are good pronounced phonemes.}
\\
Data augmentation techniques are proven to be quite effective in MD. Techniques involved (\textit{i}) modifying the canonical text with mismatched character pair keeping its original word-level speech intact \cite{non_ASR_3}; (\textit{ii}) using Text-To-Speech to synthesize `incorrect stress' samples along with modified lexical stress \cite{l2_GEN_SSL}; (\textit{iii}) using a mixup technique, in feature space -- based on phone-level GOP pool to design word level training data \cite{mixup}; and (\textit{iv}) using error distance of the clustered self supervised learning model embedding to replace the phoneme sound with similar unit \cite{ssl, ssl_2}. 

These techniques are heavily motivated by the `Cut/Paste' approach. However, the L2 learners' pronunciation is highly affected by their native phoneme set \cite{L2_effect,l2_arctic}, which makes such mixup augmentation methods less efficient.

In this study, we introduce SpeechBlender – a novel fine-grained data augmentation pipeline for generating mispronunciation errors using only good pronunciation units in low-resource settings. Our method can generate MD training data from any native/non-native speech dataset. Our framework linearly interpolates raw input signals utilizing different masks and mixing factors, allowing modification of different regions of a target phoneme\footnote{Presence of a zero/few pronunciation mistakes in the training data.}. While the proposed technique is studied for mispronunciation detection, the SpeechBlender can be broadly applicable as an augmentation technique for most sequence labeling such as accented and code-switching ASR or classification task with audio/speech as input.
Our contributions are:
\vspace{-0.1cm}
\setlength\itemsep{-0.5em}
\begin{enumerate}[label=(\alph*)]
\item Introduce the SpeechBlender, a fine-grained data augmentation pipeline capable of smoothly modifying regions of phonetic unit to generate erroneous pronunciation; 
\item Benchmark with text-based and GOP-based augmentation for Speechocean762 and AraVoiceL2 datasets;
\item Analyse different masks in the SpeechBlender and benchmark with the cut/paste signal approach;
\item Release the code for the SpeechBlender augmentation.\footnote{The code will be shared upon paper acceptance.}
\end{enumerate}

\begin{figure}[!ht]
\centering
\includegraphics[width=\linewidth]{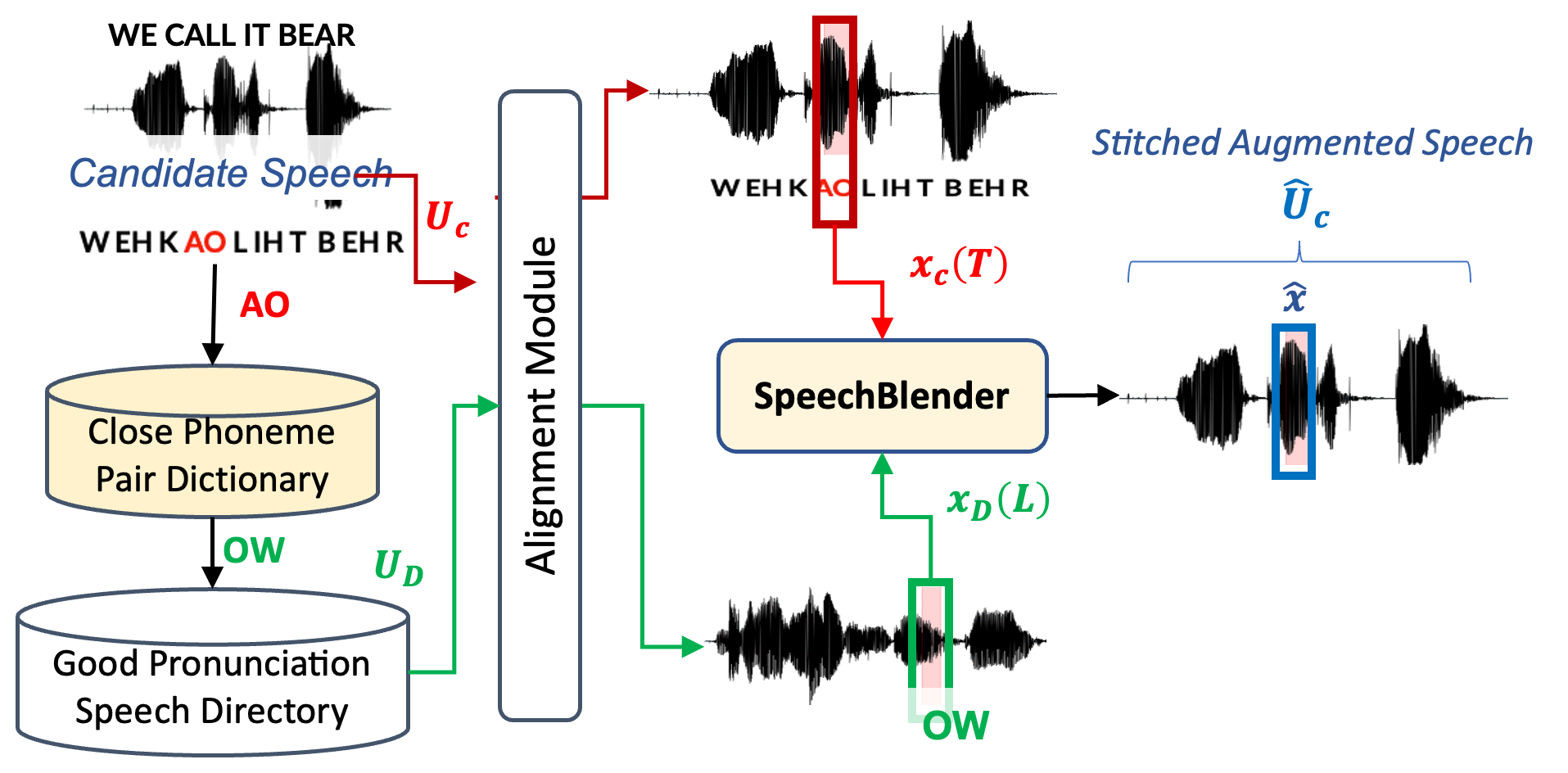}
\caption{Overview of the SpeechBlender, a fine-grained data augmentation pipeline, for zero-shot mispronunciation class generation. $U_C$: input speech utterance for candidate phoneme; $U_D$: selected speech utterance for donor phoneme; $x_C(T)$, $x_D(T)$ candidate and donor phoneme frames of length $T$ and $L$ respectively; $\hat{x}$ is the resultant augmented phoneme.}
\label{fig:data_aug_pipeline}
\end{figure}

\section{Methodology} 
\label{sec:method}

Figure \ref{fig:data_aug_pipeline} shows an overview of the SpeechBlender. $U_C$ represents an input utterance consisting of a sequence of canonical phonemes. For \textbf{candidates phonemes} selected from $U_C$,
we consult a \textbf{close phoneme-pair dictionary} and select a \textbf{donor phoneme}. We select another utterance, $U_D$, that features the donor phoneme being pronounced accurately. This selected utterance, $U_D$, serves as a source signal for the donor phoneme.

We obtain the alignments of the utterances $U_C$ and $U_D$ using a pre-trained ASR, and get the corresponding phoneme time boundaries. Using the start and end boundary of the candidate, $x_C$,  and donor, $x_D$, phoneme, we select the speech frames and apply augmentation with the SpeechBlender. We then replace the candidate phoneme $x_C$ with the augmented phoneme frames $\hat{x}$ in the candidate utterance $U_C$.

\subsection{Close Phoneme Pair Dictionary}
We randomly select a donor phoneme from a list of confused phonemes associated with the candidate phoneme. For English, the dictionary is constructed using L2-ARCTIC \cite{l2_arctic} confusing pairs matrix, and using the Speechocean762 \cite{speechocean} training set, and for Arabic, a similar dictionary was constructed using \cite{algabri2022mispronunciation} confusion pairs matrix. The entries are selected by analysing the confused phoneme pairs of speaker's pronunciation using ASR errors like substitution along with human annotation labels. Some examples of close phoneme pairs are presented in Table \ref{tab:examples_phoneme_pairs}.

\begin{table}[!ht]
\centering
\scalebox{0.9}{
\begin{tabular}{c} 
 \hline
 Close Phoneme Pairs (Confused \%) \\
 \hline\hline
 $SH \Leftrightarrow S$ (76\%), $V \Leftrightarrow F$ (44\%) ,  $NG \Leftrightarrow N$ (43\%), \\  $IY \Leftrightarrow IH$ (33\%),
 $Z \Leftrightarrow S$ (77\%) \\
 \hline
\end{tabular}
}
\caption{Examples of entries in Close Phoneme Pair Dictionary, with their \% of confusion.}
\label{tab:examples_phoneme_pairs}
\vspace{-0.6cm}
\end{table}

\subsection{Pre-trained ASR and Alignment Module}
\label{ssec:asr}
We align the acoustic frames with the corresponding reference phoneme to obtain the start and end timestamps. For the force alignment, 
we employ the Hidden Markov Model-Time Delay Neural Network (HMM- TDNN)\footnote{https://kaldi-asr.org/models/m13} trained on Librispeech corpus as the acoustic model for SpeechOcean762 dataset alignment, and we use the MGB-2 \cite{ali2016mgb} for AraVoiceL2 alignment. Both acoustic models are trained using Kaldi \cite{kaldi}.


\begin{figure*} [!ht]
\centering
\vspace{-0.4cm}
\includegraphics[width=0.8\textwidth]{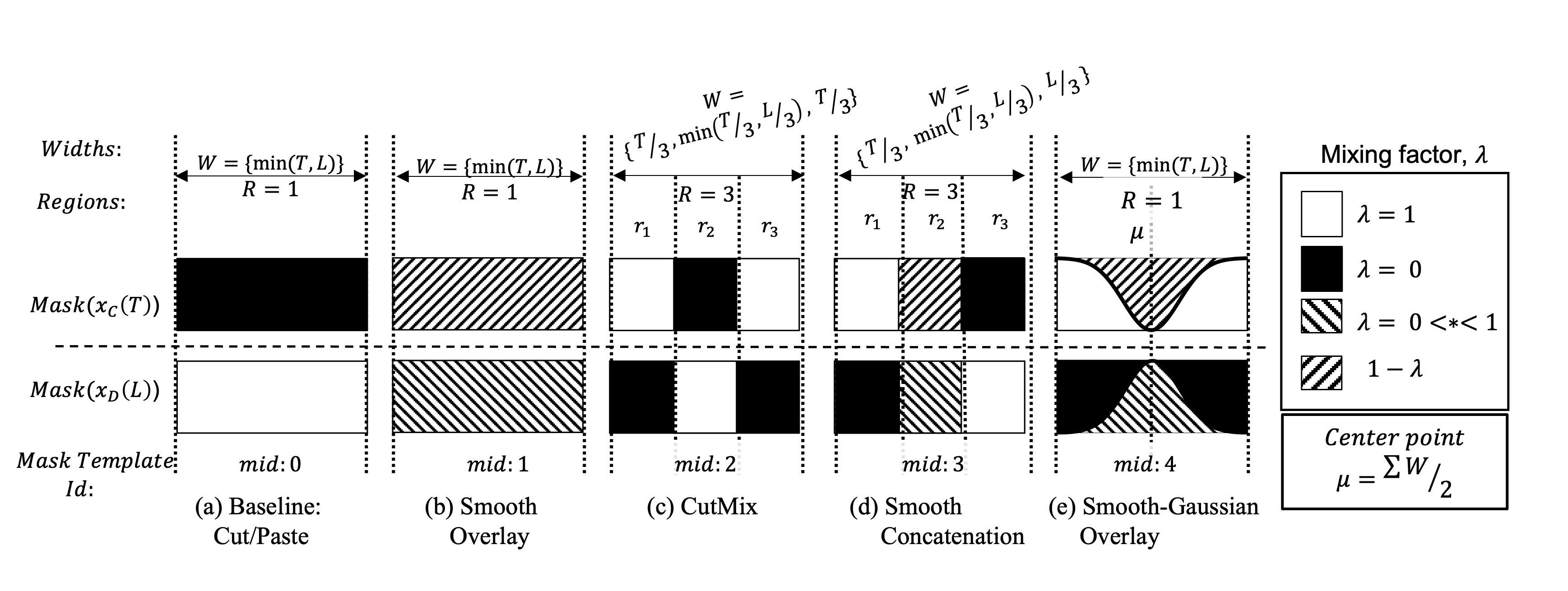}
\vspace{-0.5cm}
\caption{Comparison of the different masks used in the SpeechBlender augmentation framework. $x_C$ represent candidate phoneme (with length $T$) in the good data, augmented with the donor phoneme $x_D$ (length =$L$). 
$\mu$: the center-point; $R$: the number of regions the mask is divided into; $W$: list of widths of the regions; $\lambda$: the mixing factor in each region assigned in different augmentation scenarios given in (a) - (e). }
\label{fig:mask_types}
\vspace{-0.2cm}
\end{figure*}

\begin{algorithm}[h]
\caption{SpeechBlender function}
\SetAlgoLined
\KwIn{Input phoneme frame sequences $x_C$ and $x_D$ where $x_C = \{f^1_C \cdots f^T_C\}$ and $x_D = \{f^1_D \cdots f^L_D\}$, where ($T$, $L$) are total frames in ($x_C$, $x_D$) respectively. $y_C$ represent the original output score assigned to the good phoneme.}
\KwOut{Augmented frame output sequences $\hat{x}$ and the corresponding label $\hat{y}$.}

\SetKwFunction{FMain}{Blender}
    \SetKwProg{Fn}{Function}{:}{}
    \Fn{\FMain{$x_C, x_D, y_C$}}{
        $\theta_{\hat{x}} \longleftarrow [ ]$ \Comment{Store augmented output per region}
        
        $\theta_{\hat{y}} \longleftarrow [ ]$ \Comment{Store corresponding partial output score}
        
        $mid \longleftarrow rand([1,4])$ \Comment{Mask template id}
        
        $\rho \longleftarrow get\_property(mid, len(x_C), len(x_D)) $ \Comment{Mask property}
        
        $ W, \Lambda \longleftarrow generate\_mask(\rho)$ 

        \Comment{$W$: list of regional mask width, $\Lambda$: list of regional mixing factor values.}\


        \ForEach{$ w \in W$ and $ \lambda  \in \Lambda $}
        { 
        
        $x{_C^w}$$, x{_D^w}$ $\longleftarrow$ $x{_C}[w]$, $x{_D}[w]$ 
        
        $\theta_{\hat{x}} \longleftarrow$ $\lambda \cdot x{_C^w}$ + $(1-\lambda ) \cdot x{_D^w}$ 
        
        $\theta_{\hat{y}} \longleftarrow \lambda \cdot y{_C}$
        }
        $\hat{x} \longleftarrow  Concat(\theta_{\hat{x}})$ 
        
        $\hat{y} \longleftarrow   \left \lfloor \sum\theta_{\hat{y}}/len(\Lambda)\right \rfloor$

        \textbf{return} $\hat{x} , \hat{y} $ 
}
\label{algo}
\end{algorithm}
\vspace{-0.2cm}

\subsection{SpeechBlender}
Given the input frames of \textit{good pronounced} phoneme candidate and donor phoneme, $x_C$ and $x_D$, of length $T$ and $L$ respectively, our approach outputs the augmented phoneme $\hat{x}$ along with its corresponding output label $\hat{y}$. We first normalize the energy of $x_D$ to match $x_C$. Subsequently, we apply the SpeechBlender function, shown in Algorithm \ref{algo}, with a selected mask templates.   

\paragraph*{Mask Templates:}  
We randomly select a mask template id, $mid \in [1-4]$. We then generate the property $\rho$ of the mask based on the $mid$, length of the phonemes $T$ and $L$. The property $\rho \in \{\mu, R, W, \Lambda\}$ is parameterized by the features of the mask including: \textit{(a)} the center-point $\mu = \frac{min(T,L)}{2}$; \textit{(b)} the number of regions $R\in \{1, 3\}$ the mask is divided into; \textit{(c)} along with a list containing all widths of the regions, $W$; and \textit{(d)} a list $\Lambda$, with its corresponding mixing factor with each value $\lambda \in [0-1]$. The width of each region is determined by the minimum frame length of inputs when both the input pair are considered with $\lambda > 0$. A $\lambda = 0$ for a signal represent a mixing weight of zero, meaning the full signal is blocked out.
Details of different mask templates along with their property, $\rho$, are presented in Figure \ref{fig:mask_types}.

\paragraph*{Blending the Speech signals:} 
The Given property $\rho$ of the mask, we generated the mask with its $W$ list and its corresponding $\Lambda$ list containing region-based $\lambda$ values. 
Then for each regions in the mask, we multiply the $x{_C^w}$ and $x{_D^w}$ with $\lambda$ and $1- \lambda$ respectively (shown in Algorithm \ref{algo}) to create the regional augmented data $\theta_{\hat{x}}(.)$. We create the new augmented phoneme, $\hat{x}$ by concatenating all the $\theta_{\hat{x}}$ regional data.
The label $\hat{y} \in \{0,1\}$ of the augmented phoneme $\hat{x}$ is generated by averaging the regional label ($\theta_{\hat{y}} = \lambda \cdot y$). The $y$ is the value assigned to a good pronunciation of a phoneme.\footnote{These values are determined based on the original data annotation scheme of the respective dataset. For this study, $y=2$.} The augmented output $\hat{y}=0$ represent the mispronounced or missing phoneme; whereas $\hat{y}=1$ represent the accented, aberrant phoneme unit. In our experiment, the regional augmented output is assigned $\theta_{\hat{y}} = 0$ if the mixing factor $\lambda<0.25$, otherwise it is assigned $\theta_{\hat{y}} = 1$.

\subsection{Baseline Augmentation Methods}
\paragraph*{Text-based Augmentation}
We increase the number of error phoneme by altering the phoneme embedding from the canonical phoneme sequence. The candidate phoneme was replaced with a selected donor phoneme, using the same `close phoneme pair dictionary', when creating the canonical phoneme embedding space. An output label $\hat{y}=0$ is assigned when the candidate phoneme is swapped with a distant donor phoneme, for e.g., $SH \Leftrightarrow V$; whereas $\hat{y}=1$ is assigned for a close phoneme exchange such as $SH \Leftrightarrow S$.

\paragraph*{GOP-based Augmentation}
We first create a bag of GOP feature embedding for all good pronounced phonemes and grouped them by phoneme pairs using the same `close phoneme pair dictionary'. We then replaced the candidate phoneme, in the utterance with a donor phoneme GOP embedding randomly selected from the bag. We follow the same scoring scheme as text-augmentation for $\hat{y}$ labels.

\section{Experiments} 
\label{sec:exp}


\subsection{Speech Corpus}
Table \ref{tab:output_defination} shows details about the datasets used in our study. For English, we opted for the Speechocean$762$ \cite{speechocean}, 
which consists of $5,000$ English utterances from $250$ non-native speakers. Five experts annotate each phoneme. 
For Arabic, we use an in-house corpus, AraVoiceL2, which is comprised of $5.5$ hours of data recorded by $11$ non-native Arabic speakers. Each speaker recorded a fixed list of $642$ words and short sentences, making for a total of $7,062$ recordings. We adapted the guideline from Speechocean$762$ dataset.\footnote{Citation to actual annotation guideline removed for anonymity.} We split AraVoiceL2 data, $6$ speakers for training and $5$ speakers for testing. For the study, we designed the MD model as classification task.

\begin{table*}[!ht]
\centering
\scalebox{1.0}{
\begin{tabular}{c|c c | c c | c c } 
 \hline
  &   \multicolumn{2}{c}{\textbf{Speechocean762}} & \multicolumn{2}{c}{\textbf{AraVoiceL2 phoneme}}                                         &  \multicolumn{2}{c}{\textbf{AraVoiceL2 grapheme}}   \\
 Label  &  Training & Testing & Training & Testing & Training & Testing \\
 \hline\hline
 0     & $1,991$ & $1,339$ & $97$ & $94$ &  $95$ & $89$ \\ 
 \hline
 1   & $1,967$ & $1,828$ & $707$ & $618$ & $590$ & $483$ \\
 \hline
 2   & $43,118$ &  $44,079$ & $28,270$  &  $23,008$ & $20,255$  &  $16,511$ \\
 \hline
\end{tabular}
}

\caption{Phoneme-level statistics for Speechocean762 corpus, and phoneme, grapheme-level statistics for AraVoiceL2 corpus, along with the score definition used in the study. Label \textit{0}: mispronounced or missing phoneme, Label \textit{1}: accented pronunciation and Label \textit{2}: is for good pronunciation.}
\label{tab:output_defination}
\vspace{-0.7cm}
\end{table*}

\subsection{Mispronunciation Detection Model}


We first extract the frame-level phonetic features using pre-trained ASR, described in Section \ref{ssec:asr}. Using the acoustic model, we calculated the 84-dimensional GOP features for 42 pure phonemes. 
Each phone GOP representation contain two different features: \textit{(i)} log phone posterior (LPP) \cite{transfer_} of a phone $p$ of length $T$ frames with per frame observation $o_t$ see Equation \ref{eq:lpp}; 
and \textit{(ii)} the log posterior ratio (LPR) \cite{transfer_} of phone $p_{j}$ with respect to phone $p_i$, as given in Equation \ref{eq:lpr}.

\noindent These 84-dimensional features are then passed through a feed-forward layer to project the GOP embeddings in 24 dimensional feature space.  
Simultaneously, we inject the textual-content information using the one-hot vector representation of the canonical phoneme transcription. These vector representations are then projected to 24-dimensional phoneme embedding and are then added to the GOP representations.

\begin{equation}
\label{eq:lpp}
\begin{small}
LPP(p) = \frac{1}{T} \sum log\ p(p|o_t)
\end{small}
\end{equation}
\begin{equation}
\label{eq:lpr}
LPR(p_j|p_i) = log\ p(p_j|o_t) - log\ p(p_i|o_t)
\end{equation}

\noindent Consequently, we pass the final representation to the encoder network. As our encoder, we opt for a long short-term memory (LSTM)\footnote{Additional models such as SVR\cite{speechocean}, MLP are tested with SpeechBlender data augmentation. As the results follow the same trend , for brevity we are reporting the results using LSTM only.}
network due to the model's strength in capturing preceding contextual information, and its reduced computational cost without losing much on the model accuracy. For further comparison, we implement the architecture with similar configuration as \cite{JIM} with 3 stacked LSTM layers and an output embedding dimension of 24. We then pass the embedding per phoneme to the output layer for the regression task.

\noindent\textbf{Model Parameters}: We train the mispronunciation detection model using an Adam optimizer for $50$ epochs with a starting learning rate of $1e-3$ and a batch size of $32$. The learning rate is then cut in to its half in every 3 epochs, after 10 initial epochs. 
Regarding the loss function, it is adapted to suit the specific annotations dataset. Specifically, for the English dataset, which involves a continuous task and scores ranging from 0 to 2, we have utilized the mean squared error (MSE). In contrast, for the Arabic dataset, which involves a classification task, we adopted cross entropy.

\subsection{Evaluation Measures}
We measure the proposed models performance using standard statistical metric: 
We report MSE and PCC for the English dataset, Precision (PR), Recall (RE), and F1-score for the Arabic Dataset. We use PCC, F1-score for ranking as they are more representative of the performance in such an imbalanced scenario.

\section{Results and Discussion}

\subsection{Comparison with Baselines}
\paragraph*{Speechocean762:} 
We benchmark our results in Table \ref{tab:results_1}, the PCC results indicate the importance of augmentation in general. 
The results indicates that both the Text-Augmentation and GOP-augmentation outperforms the model trained with only Speechocean762 by $1\%$, while SpeechBlender shows a performance improvement of $5\%$.
Our novel augmentation technique also surpassed the performance of GOP-based LSTM \cite{JIM} and the
state-of-the-art GOP-based Transformer model \cite{JIM} by $4$\% and $2\%$, respectively, using only phoneme-level information only in contrast to their multi-task setup. 




\begin{table} [!ht]
\centering
\arrayrulecolor{black}
\begin{tabular}{!{\color[rgb]{0,0,0}} l|l|l} 
\arrayrulecolor{black}\hline
\textbf{Training Dataset}                      & \textbf{PCC} & \textbf{MSE}                                           \\ 
\arrayrulecolor{black}\hline\hline
\multicolumn{3}{c}{\textbf{Baselines}}                                                                                  \\ 
\hline\hline
\textit{Speechocean762-Train}                     &     0.58         &     0.090 \\   
\hline 
\textit{+ Multi-task Learning (w. LSTM) \cite{JIM}}    &     0.59         &     0.089                                                   \\ 
\hline 
\textit{+ Multi-task Learning (w. Transformer) \cite{JIM}}    &     0.61         &     0.085                                                   \\ 
\hline\hline
\multicolumn{3}{c}{\textbf{ Augmentation Baselines}}                                                                                  \\ 
\hline\hline
\textit{+ Text-augmentation}            &       0.59       &                  0.092                                     \\ 
\hline
\textit{+ GOP-augmentation}             &     0.59         &                    0.090                                    \\ 
\hline\hline
\multicolumn{3}{c}{\textbf{Our Proposed Augmentation Method}}                                                                              \\ 
\hline\hline
\textit{+ SpeechBlender-Aug} &         \textbf{0.63}     &                 \textbf{0.085} \\ \hline

\end{tabular}

\caption{Reported phoneme-level PCC and MSE on Speechocean762 testset}
\vspace{-0.7cm}
\label{tab:results_1}
\end{table}

\paragraph*{AraVoiceL2:} We report the effectiveness of the SpeechBlender framework on the Arabic AraVoiceL2 testset. Arabic is a consonantal language with high variability in the usage of vowels -- that changes based on grammar, dialectal properties and its intend meanings. For a generalization over dialects and to overcome the aforementioned challenges, we focus our analysis solely on the grapheme level for this study. Our reported result (in Table \ref{tab:results_5}) shows an improvement of 4.63\% F1-score on AraVoiceL2 testset over the baseline Arabic model. 
We also observed a gain of 1.98\% and 1.44\% over the Text-augmentation and GOP-augmentation.
Evidently, the results shows the efficacy of different augmentation and specially SpeechBlender for generating erroneous pronunciation in a zero/few-shot setting. 



\begin{table} [!ht]
\centering
\arrayrulecolor{black}
\scalebox{1.0}{
\begin{tabular}{!{\color[rgb]{0,0,0}} l|l|l|l} 
\arrayrulecolor{black}\hline
\textbf{Training Dataset}                      & \textbf{Precision} & \textbf{Recall} & \textbf{F1-score}                 \\ 
\arrayrulecolor{black}\hline\hline
\multicolumn{4}{c}{\textbf{Baselines}}                                                                                  \\ 
\hline\hline
\textit{AraVoiceL2-Train}                     &     61.72\%          &     53.78\% &  55.37\% \\   
\hline\hline
\multicolumn{4}{c}{\textbf{Augmentation Baselines}}                                                            \\ 
\hline\hline
\textit{+ Text-augmentation}            &       60.08\%      &       56.57\% & 58.08\%         \\ 
\hline
\textit{+ GOP-augmentation}             &    59.26\%      &       57.98\% & 58.56\%                                 \\ 
\hline\hline
\multicolumn{4}{c}{\textbf{Our Proposed Augmentation Method}}                                                                              \\ 
\hline\hline
\textit{+ SpeechBlender-Aug} &      60.33\%      &       \textbf{59.69}\% & \textbf{60.00}\%      \\ \hline

\end{tabular}
}
\caption{Reported Grapheme-Level Precision, Recall and F1-score on AraVoiceL2 testset}
\vspace{-0.6cm}
\label{tab:results_5}
\end{table}

\subsection{Effectiveness of the Masks } We assess the effectiveness of each standalone mask incorporated in the SpeechBlender for creating different mispronunciation instances. We reported our analysis with Speechocean762 data for brevity. Our results, in Table \ref{tab:resuts_2}, shows that the smooth blending masks (Smooth Overlay and Smooth-Gaussian Overlay) create more practical variation in mispronunciation and accented phonemes than their counter-part. When compared directly with the `Cut/Paste' augmentation mask, our blending mask also shows its efficacy. Smooth Overlay masking shows a relative improvement of 4\% compared to no-augmentation scenario. We observed a similar pattern with AraVoiceL2 dataset.

\begin{table} [!ht]

\centering

\scalebox{1.0}{
\arrayrulecolor{black}
\begin{tabular}{!{\color[rgb]{0,0,0}} l|c|c} 
\arrayrulecolor{black}\hline
\textbf{Augmentation Mask}                      & \textbf{PCC} & \textbf{MSE}                                           \\ 
\arrayrulecolor{black}\hline\hline
\multicolumn{3}{c}{ \textit{Mask creating accented phoneme class, $\hat{y}=1$}}                                                                                  \\ 
\hline\hline

\textit{Smooth Overlay ($\lambda = 0.5$)}               &      \textbf{0.62}       &               \textbf{0.086}                                        \\ 
\hline
\textit{CutMix}                       &        0.59      &                   0.088                                     \\ 
\hline
\textit{Smooth Concatenation}         &      0.59        &              0.089                                          \\ 
\hline
\textit{Smooth-Gausian Overlay}       &      0.60        &   0.088\\
\hline
\textit{Smooth Overlay ($\lambda = 0.6$)}       &      0.61        &    0.087\\
\hline\hline
\multicolumn{3}{c}{ \textit{Mask creating mispronounced phoneme class, $\hat{y}=0$}}                                                                                  \\ 
\hline\hline
\textit{Cut/Paste}                         &         0.58    &                0.104                                        \\ 
\arrayrulecolor{black}\hline
\textit{Smooth Overlay ($\lambda = 0.1$)}               &      0.56       &                0.096                                    \\ 
\hline
\textit{Smooth Overlay ($\lambda = 0.2$)}       &      \textbf{0.58}        &    \textbf{0.092}\\\hline

\end{tabular}
}
\caption{Reported phoneme-level PCC and MSE on Speechocean762 test for different augmentation mask}

\vspace{-0.6cm}
\label{tab:resuts_2}
\end{table}



\section{Conclusion} \label{sec:conclusion}

In this paper, we introduce a fine-grained data augmentation technique, SpeechBlender, and show its efficacy in generating new error classes for a phoneme-level and grapheme-level pronunciation quality detection task. The SpeechBlender linearly interpolate raw input signals using different blending masks like `\textit{SmoothOverlay}' and mixing factors $\lambda$. Varieties of masks included in the proposed method allows smooth modification in different regions of the phoneme.  
With SpeechBlender, we observe improved performance compared to \textit{(i)} no-augmentation, \textit{(ii)} multi-task learning scenario, \textit{(iii)} text-augmentation, and \textit{(iv)} GOP-augmentation setting. We noticed the \textit{`Cut/Paste'} method be it in text-, feature- or raw-audio domain, is not as efficient as the proposed interpolation technique.

Moreover, our results suggest that with blending the signals, we can generate new positive error classes instead of reusing samples from the data itself in zero-shot setting. 
The different choices of masks allow variations in transition and style of errors. In future, we will further explore the capability of SpeechBlender for generating accented/erroneous phoneme candidate for other speech tasks like ASR, while exploring multiple granularities and supra-segmental features.

\bibliographystyle{IEEEtran}
\bibliography{mdd}

\end{document}